\documentclass[10pt]{iopart}
\usepackage{iopams,color}

\begin{document}
\title[{Heisenberg-style bounds for arbitrary estimates of shift parameters}]{Heisenberg-style bounds for arbitrary estimates of shift parameters including prior information}
\author{Michael J W Hall and Howard M Wiseman}
\address{Centre for Quantum Computation and Communication Technology (Australian Research Council), Centre for Quantum Dynamics, Griffith University, Brisbane, QLD 4111, Australia}



\begin{abstract}
A rigorous lower bound is obtained for the average resolution of any estimate of a shift parameter, such as an optical phase shift or a spatial translation.  The bound has the asymptotic form $k_I/\langle2|G|\rangle$ where $G$ is the generator of the shift (with an arbitrary discrete or continuous spectrum), and hence establishes a universally applicable  bound of the same form as the 
usual Heisenberg limit.  The scaling constant $k_I$ depends on prior information about the shift parameter.   For example,  in   phase sensing regimes, where the  phase shift  is confined   to some small interval of length $L$, the relative  resolution  $\delta\hat{\Phi}/L$ has the strict lower bound $(2\pi e^3)^{-1/2}/\langle 2m| G_1 |+1\rangle$, where $m$ is the number of probes,  each with generator $G_1$,  and entangling joint measurements are permitted. Generalisations  using  other resource measures and  including   noise are briefly discussed.  The results rely on the derivation of general entropic uncertainty relations for continuous observables, which are of   interest  in their own right.
\end{abstract}


\section{Introduction}

In many measurement scenarios, an environmental variable acts to translate or shift a property such as the optical phase or position of a probe state. Accurate estimation of the shift parameter allows a correspondingly accurate measurement of the environmental variable.  For example, interferometric measurements of  quantities such as temperature, strain and gravitational wave amplitudes rely on estimation of an optical phase shift.  An important aim of quantum metrology is to determine the fundamental  bounds   on the resolution of such estimates  \cite{glm2004,WisMil10}.

 One useful tool in this respect is the quantum Cramer-Rao inequality \cite{helstrom,holevo}, which can be used to obtain bounds on the resolution of a shift parameter  in terms of  the variance of the operator that generates the shift \cite{WisMil10,helstrom, holevo,cr,paris}.  For example, the root mean square error of any unbiased estimate $\hat{X}$ of a shift parameter $X$, for any fixed value $X=x_0$, satisfies \cite{cr,paris}
\begin{equation} \label{cr} \delta\hat{X}_{x_0} =\langle (\hat{X}-x_0)^2\rangle^{1/2}\geq {1}/(2\sqrt{m}{\Delta G}) , 
\end{equation}
where $G$ is the shift generator, $\Delta G$ is the uncertainty of $G$ for the probe state, and $m$ is the number of (independently measured) copies of the probe state. 

One can also use the Cramer-Rao inequality to derive resolution bounds in term of other quantities, such as the maximum and minimum eigenvalues of $G$ when they exist \cite{glm06,boixo}, or the mixedness of the probe state \cite{modi}.  While such bounds are universally valid, scientists  often  wish to obtain bounds in terms of other resources, such as the average energy or mean photon number of the probe state. 
But, for example, if $G=N$ is the photon number of a single mode field, then $\Delta N$ can be arbitrarily large  or small relative to the average photon number $\langle N\rangle$.  A simple example which proves this point   is a probe state with number distribution $p_0=1-w$, $p_n=w$ for some $n\neq 0$, and vanishing otherwise. Then  $\langle N\rangle/\Delta N = [ (1-w)/w]^{1/2}$  ranges over $(0,\infty)$  as $w$ ranges over $(0,1)$.\footnote[1]{Moreover, if one chooses $w=n^{-3/2}$, then the value of $\Delta N$ diverges with increasing $n$, while $\langle N\rangle$  becomes arbitrarily small.}  Thus, the lower bound (\ref{cr}) does not limit phase resolution in terms of average photon number.   This illustrates that  the quantum Cramer-Rao inequality  cannot always be used to derive bounds in terms of  the resources  of interest  required to achieve a given resolution, 
 in particular when the eigenvalue range is unbounded.

An alternate tool for bounding resolution  is the Heisenberg limit \cite{heis}.  While less developed than the quantum Cramer-Rao inequality, it may be heuristically characterised as an asymptotic lower bound on measurement resolution that scales inversely with the number of resources available  and that is achievable, up to a constant factor   \cite{summy,jmodopt,luis,WisKil98,BWZ99,berryreview,zwierz}.  
These references use various ways to quantify meausurement resolution and number of resources, but 
at this point it suffices to quote the earliest result (obtained numerically) \cite{summy}: 
 the root mean square error of a canonical phase estimate on a single mode field, for any applied phase shift $\phi_0$,  can asymptotically scale no better than 
 \[  \delta \hat{\Phi}_{\phi_0} \gtrsim {1.38}/{\langle N\rangle}. \]
This scaling with  the average photon number  $\langle N\rangle$, rather than with $\Delta N$  as in (\ref{cr}),  has the advantage of providing a necessary condition on the energy resources required for a given phase resolution.

 Recently, progress  has been made in generalising the Heisenberg limit  to obtain {\it non}-asymptotic resolution bounds for  {\it arbitrary} estimates of shift parameters.  These may be called Heisenberg-style bounds.  First, for a shift generator $G$ with a discrete spectrum and finite lowest eigenvalue $g_{\rm min}$,  
Giovannetti \etal have bounded the root mean square error of any estimate, when averaged over two fixed values of the shift parameter, by $0.076/\langle G-g_{\rm min}\rangle$  ---  providing that the fixed values and errors satisfy a particular constraint \cite{glm}.  Second, when $G$ is further restricted to integer eigenvalues, Hall \etal have obtained the constraint-free bound of $0.559/\langle G-g_{\rm min}+1\rangle$, for the root mean square deviation of any estimate, when uniformly averaged over {\it all} values of the shift parameter \cite{prl}. 

Despite this progress,  none of the above results characterise the overall performance of a given estimate in the important case that prior information is available about the value of the shift parameter.  For example, in a  {\em phase sensing} regime  \cite{sensing},   such as gravitational wave detection, the value of an applied phase shift is  {\it a priori} known to lie within some small interval about zero. Hence, only the performance of the estimate over the interval is of interest  ---  it is irrelevant how well or how badly the estimate may perform outside this interval.  Indeed, in a stimulating paper, Rivas and Luis have recently proposed a phase estimation scheme that improves on the scalings of the above Heisenberg bounds for small phase shifts \cite{rivas}. 

It is  therefore of great   interest to determine fundamental bounds when prior information is available to be exploited.  For example, the quantum van Trees inequality generalises the quantum Cramer-Rao inequality, to bound the mean square error averaged over the  prior probability density $q(x)$ of the shift parameter \cite{paris,gill}, leading to $\Delta G$ in (\ref{cr}) being correspondingly replaced by $[{\rm Var\,}G +F_q]^{1/2}$, where $F_q=\int q(x)\,(d\ln q(x)/dx)^2$ is the Fisher information of $q(x)$.  Note however that, for estimates confined to some bounded interval with $q(x)$ uniform over this interval, $F_q$ vanishes and no improvement is obtained over the lower bound in (\ref{cr}).  

Tsang has very recently  obtained the first results  exploiting prior information in the context of Heisenberg-style bounds \cite{tsang}.  For example, he has shown that, for discrete shift generators and a prior density $q(x)$ uniform over any interval of length $L$, the root mean square error of any estimate is bounded by a function which asymptotically approaches $0.154/\langle G-g_{\rm min}\rangle$, under a constraint that $L$ is sufficiently large (see also Ref.~\cite{gm}).  

The central result of the present paper is a general  Heisenberg-style lower  bound that takes arbitrary prior information into account.  It is constraint-free, does not require the generator to be discrete, nor to have a finite lowest eigenvalue.   In particular,  
we show that the average mean square deviation, $(\delta \hat{X})^2 = \langle (\hat{X}-X)^2\rangle$, of any estimate $\hat{X}$ of some shift parameter $X$, over any prior distribution of $X$,  satisfies
\begin{equation} \label{gen}
\delta \hat{X} \geq \frac{k_I}{c+2\Delta^{-1}\langle \,|G-g|\, \rangle } ~.
\end{equation}
Here $k_I$ depends on the prior information available about the shift parameter, $G$ is the shift generator, and $g$ is an arbitrary eigenvalue of $G$.  For continuous spectra, $c=0$ and $\Delta=1$, while for discrete spectra $c=1$ and $\Delta$ is the minimal spectral gap of $G$.  Further, if the spectrum of $G$ has a smallest value, $g_{\rm min}$, then the factor of $2$ in the denominator can be removed for the choice $g=g_{\rm min}$.    

The scaling constant $k_I$ in (\ref{gen}) depends on the prior distribution of the shift parameter, and can in principle be arbitrarily small for a   sufficiently narrow prior distribution.  In particular, for any generator with a discrete spectrum,  if the shift parameter $X$ is known {\it a priori} to be randomly distributed over an interval of length $L$, it will be shown that 
 $k_I \geq L (2\pi e^3)^{-1/2}$.  Thus there is a fundamental `relative resolution' bound for how well the shift can be resolved, relative to the size of the interval to which it is confined. Moreover, with the additional assumption that the state is 
a number $m$ of identical copies of some probe state, each with  discrete  generator  $G$,  it will be shown that inequality (\ref{gen}) implies that 
\begin{equation}\label{gendis}
\frac{\delta\hat{X}}{L} \geq \frac{(2\pi e^3)^{-1/2}}{1+ 2m\Delta^{-1}\langle \,| G  -g|\, \rangle } ,
\end{equation}
with a similar bound conjectured when $G$ has a continuous spectrum.  Note that this inequality allows for arbitrary measurements over the $m$ probes, even entangling joint measurements,  unlike bounds obtained from the Cramer-Rao inequality, which assumes a fixed measurement. 
The scaling of (\ref{gendis}) with $m^{-1}$ contrasts to the $m^{-1/2}$ scaling in (\ref{cr}). While neither bound is necessarily achievable 
for a given probe state and a given $m$, the scaling as $m^{-1/2}$ in (\ref{cr}) is as expected from elementary statistics. Hence 
we do not expect that our bounds would be tight for a fixed probe state and large $m$.  

Similarly to (\ref{gen}), if the spectrum of $G$ has a smallest value $g_{\rm min}$ then the factor of $2$ in the denominator of (\ref{gendis}) can be removed for $g=g_{\rm min}$.  
For example, for phase shifts generated by the photon number operator $N$ of a single mode field, the relative resolution of any estimate over an interval of length $L$ is bounded below by $(2\pi e^3)^{-1/2}/\langle 1+mN\rangle$, generalising the result of Hall \etal \cite{prl},  which was limited  to the case of a completely random phase shift, with $L=2\pi$. 

The results of the paper are obtained via the derivation of suitable entropic uncertainty relations for the shift generator $G$ and the error $\hat{X}-X$ (section 2 and appendices). Examples are given for estimates of optical phase, time, and spatial displacements, including generalisations to alternative resource measures such as the  support of the energy distribution  of the probe state (sections 3 and 4).  A further generalisation of inequalities (\ref{gen}) and (\ref{gendis}) is given which quantifies the effects of noise (section 5), followed by a brief discussion (section 6).

\section{Estimation bounds from entropic uncertainty relations}

\subsection{General estimation schemes}

Consider a general shift parameter estimation scheme, in which a probe state $\rho_0$ undergoes a shift generated by some operator $G$, to the state $\rho_x=e^{-iGx}\rho_0e^{iGx}$.  A measurement of observable $M$ on the probe then outputs some estimated value ${\hat{x}}$ of the actual shift value $x$, and is described by some positive-operator-valued measure (POVM)  ${\cal M} = \{M_{\hat{x}}\}$ \cite{WisMil10,holevo}.   
A standard notation will be used in which random variables and operators appear in upper case, and (eigen)values of these quantities appear in lower case.  Thus, the shift parameter will be denoted by $X$, and its estimate by $\hat{X}$.  

If the prior probability density of the shift parameter is denoted by $q(x)$, then the probability density of the error or deviation, $Y=\hat{X}-X$, of the estimated value from the true value, is given by
\begin{equation} \label{error}
 p_{Y}(y) = \int dx\, q(x)\, p_{\hat{X}}(x+y|x) = \int dx\,q(x)\,{\rm tr}[\rho_xM_{x+y}] ={\rm tr}[\rho_0\overline{M}_y] 
\end{equation}
where the POVM  $\overline{\cal M}\equiv\{\overline{M}_y\}$  is defined by
\begin{equation} \label{my}
\overline{M}_y := \int dx \, q(x) \,e^{iGx}M_{x+y}e^{-iGx} .
\end{equation}

A `good' estimate of the shift parameter will be one for which the error is small on average, i.e., for which $p_Y(y)$ is highly peaked about $y=0$.  This may be quantified by the mean square deviation
\begin{equation} \label{deltax}
(\delta \hat{X})^2 := \langle (\hat{X}-X)^2\rangle = \langle Y^2\rangle = \int dy \,y^2\,p_Y(y) .
\end{equation}
Note that for periodic shift parameters, such as phase, the integration may be taken over an interval centred on $y=0$ \cite{prl}. The quantity $\delta \hat{X}_{x_0}$ in (\ref{cr}) corresponds to the case $q(x)=\delta(x-x_0)$.

Equations (\ref{error})-(\ref{deltax}) generalise the case of phase estimates considered in Hall \etal \cite{prl}, where the prior density was restricted to be uniform, i.e.,  $q(x)=1/2\pi$. Under   this   restriction  the  POVM $\overline{ \cal M}$ is covariant,   with $\overline{M}_{x+y} = \,e^{-iGy}\overline{M}_{x}e^{iGy}$, 
  allowing a connection to be made between $\delta \hat{X}$ and $G$ via an entropic uncertainty relation for canonically conjugate number and phase operators \cite{prl}.
However,  that  method fails whenever $q(x)$ is non-uniform, as $\overline{ \cal M}$ is no longer covariant. Note also that $q(x)$ is necessarily non-uniform for non-periodic shift parameters.

It turns out that the key to generalising the approach of Hall \etal is the extension of existing entropic uncertainty relations for arbitrary discrete observables \cite{para, coles, berta}, to the case of continuous POVMs, such as $\overline{ \cal M}$ in equation (\ref{my}) above.  The necessary extensions are derived in appendix A.  It will now be shown how these lead to Heisenberg-style lower bounds for $\delta\hat{X}$, as per inequalities (\ref{gen}) and (\ref{gendis}). Examples and generalisations are given in sections 3-5.

\subsection{Exploiting entropic uncertainty relations}

Suppose that one has an entropic uncertainty relation for the observables $G$ and   $\overline{M}$ (the observable corresponding to the POVM $\overline{\cal M}$)  of the form
\[ H(G) + H(\overline{ M}) \geq \ln K_I \]
for some constant $K_I$.   Here $H(A)$  denotes the Shannon entropy of the measurement distribution of observable $A$, for a probe in state $\rho_0$.   Several such uncertainty relations are given in appendix A.  In general, $K_I$ depends on the prior information encoded in the prior density $q(x)$, and its form is discussed in section 2.3 below.  

From equation (\ref{error}) the statistics of $\overline{ M}$ and $Y$ are identical, 
and thus the above uncertainty relation can be rewritten as
\begin{equation} \label{hbound}
H(\hat{X} - X)  = H(Y) \geq \ln K_I -H(G) .
\end{equation}
Furthermore,  consider the variational quantity $J=H(Y)+\alpha \langle 1\rangle+\beta \langle Y^2\rangle$, where $\alpha$ and $\beta$ are Lagrange multipliers fixing the normalisation of $p_Y(y)$ and the value of $\langle Y^2\rangle$ respectively.  The variational equation $\delta J/\delta p_Y=0$ leads directly (Chapter 12 of \cite{power}) to the upper bound
$H(Y) \leq \frac{1}{2} \ln \left[ 2\pi e \langle Y^2\rangle \right]$, saturated by the Gaussian distribution $p_Y(y) = (2\pi  \langle Y^2\rangle)^{-1/2}\exp[-y^2/2\langle Y^2\rangle]$.  Combining this bound with equations (\ref{deltax}) and (\ref{hbound}) yields the lower bound
\begin{equation} \label{xbound}
\delta\hat{X} \geq \frac{K_I}{(2\pi e)^{1/2}}e^{-H(G)}
\end{equation}
for the root mean square deviation of the estimate.  

Inequalities (\ref{hbound}) and (\ref{xbound}) provide  information-theoretic bounds on the performance of the estimate, in terms of the entropy of the shift generator for the probe state. This is already useful in contexts where entropy itself can be considered as a resource.  Inequality (\ref{xbound}) is also useful for determining alternative bounds on resolution, under various constraints on $G$, as will be discussed in sections 3 and 4.

Finally, as shown in appendix B, if $g$ is an arbitrary eigenvalue of $G$, then the entropy of $G$  is bounded above by
\begin{equation} \label{hmax}
H(G) \leq 1 + \ln \left[c+ 2\Delta^{-1}\langle |G-g|  \rangle  \right] ,
\end{equation}
where the factor of $2$ can be dropped if $G$ has a minimum eigenvalue $g_{\rm min}$ and $g=g_{\rm min}$. For discrete generators,  $\Delta = \min_{g'\neq g''}|g'-g''|$  is the minimum spectral gap of $G$, and $c=1$.  For continuous generators, $\Delta=1$ and $c=0$. Substitution of (\ref{hmax}) into  (\ref{xbound}) immediately yields inequality (\ref{gen}), with
\begin{equation} \label{ki}
k_I = (2\pi e^3)^{-1/2}K_I .
\end{equation} 

Before discussing specific examples and generalisations of the generic Heisenberg bound (\ref{gen}), the dependence of the constant $K_I$ on the prior information encoded in $q(x)$ will be examined, yielding a derivation of the relative resolution bound (\ref{gendis}) for discrete generators, and an analogous conjectured bound for continuous generators.

\subsection{Dependence on prior information}

\subsubsection{Discrete generators:}

For a shift generator $G$ with a discrete spectrum, a suitable scaling constant $K_I$ follows from inequality (\ref{eurac}) of appendix A as
\begin{eqnarray*}  K_I^{-1} &=& \sup_{g, y,\psi} \langle\psi|{ \Gamma_g}\overline{M}_y{ \Gamma_g}|\psi\rangle
= \sup_{g,y,\psi} \langle \psi_g|\overline{M}_y|\psi_g\rangle\,\langle\psi|{ \Gamma_g}|\psi\rangle \\
&=& \sup_{g,y} \langle g|\overline{M}_y|g\rangle , 
\end{eqnarray*}
where ${ \Gamma_g}$ denotes the  projector  on to the unit eigenspace of eigenvalue $g$, $|\psi_g\rangle:={ \Gamma_g}|\psi\rangle/\langle\psi|{ \Gamma_g}|\psi\rangle^{1/2}$, and $|g\rangle$ denotes any normalised eigenstate of $G$.  The last equality holds since (i) $|\psi_g\rangle$ is always equal to some $|g\rangle$ by construction, and (ii) the factor $\langle\psi|{ \Gamma_g}|\psi\rangle$ is maximised when $|\psi\rangle=|\psi_g\rangle$.  
Hence, using equation (\ref{my}) for $\overline{M}_y$, one has
\begin{equation} \label{kinverse}
K_I^{-1} = \sup_{g,y} \int dx\, q(x)\,\langle g|M_{x+y}|g\rangle = \sup_{g,y} \int dx\,q(x)\, p_g(x+y) , 
\end{equation}
where $p_g(x)$ denotes the measurement distribution when the probe state is replaced by eigenstate $|g\rangle$ of $G$.  

Thus, the value of $K_I$ is determined by the maximum possible value of the convolution of the prior probability density $q(x)$ with the measurement distributions $\{p_g\}$.  
Moreover, noting that
\[ \int dx\,q(x)\, p_g(x+y) \leq  q_{\max} \int dx \, p_g(x+y) = q_{\rm max}, \]
where $q_{\rm max}$ denotes the maximum value of the prior probability density $q(x)$, one has
\begin{equation} \label{qmax} 
K_I \geq 1/q_{\rm max} .
\end{equation}
This constraint on $K_I$ leads to the universal relative resolution bound (\ref{gendis}) for discrete generators.  In particular, an estimate based on (a possibly entangling joint measurement on) $m$ copies of  a probe state corresponds to replacing $\rho_0$ by   $\otimes^m\varrho_0$  and $G$ by $G_{ \rm T}=G_1+G_2+\dots +G_m$, where ${G_j}$ refers to $G$ for the $j$-th copy.  Note that $mg$ is an eigenvalue of $G_{ \rm T}$, $|G_1+\dots+G_m-mg| \leq |G_1-g|+\dots+|G_m-g|$, and $\Delta_{ \rm T}=\Delta$.  
Hence, choosing $q(x)$ to be uniform over an interval of length $L$ and vanishing elsewhere,  $q_{\rm max}=1/L$ and inequality (\ref{gendis}) follows from relations (\ref{gen}), (\ref{ki}) and (\ref{qmax}).  

The lower bound in (\ref{qmax}) can be approached, in principle, if the measurement distribution $p_g(x)$ is sufficiently peaked around some value $x_g$, for some eigenstate $|g\rangle$ of $G$.  In particular, this allows $y$ to be chosen in (\ref{kinverse}) such that $p_g(x+y)$ is peaked around the maximum value of $q(x)$.
Thus,  the more the estimate is concentrated around some value, given a system prepared in some eigenstate $|g\rangle$ of $G$,  the closer the constant $K_I$ will be to $1/q_{\max}$,  allowing the possibility of of approaching the lower bound in equation (\ref{gendis}).  This possibility is further discussed in section 3.1.  

 Note finally that the bound in (\ref{qmax}) is only useful when the prior probability density $q(x)$ is not infinitely peaked.  Although $q_{\rm max}$ will be finite for any physical prior distribution, it is of interest to find stronger bounds not subject to this limitation.  For example, note that any probe state corresponding to an eigenstate $|g\rangle$ of $G$ is invariant under shifts generated by $G$, implying no corresponding estimate can improve on prior knowledge about the shift parameter.  It is therefore natural to define an estimate to be `ignorance respecting' if the measurement distribution for any eigenstate of $G$,  $p_g(x)$, is not any better concentrated than the prior probability density $q(x)$, in the standard sense that $p_g$ majorises $q$ \cite{major}. This implies in particular that $\int dx\, F(p_g(x))\leq \int dx\,F(q(x))$ for any continuous convex function $F$ \cite{major}.  Choosing $F(z)=z^2$, and writing $C_r:=\int dx\,r(x)^2$ for probability density $r$, it  follows via equation (\ref{kinverse}) and the Schwarz inequality that 
\[ K_I^{-1} = \sup_{g,y} \int dx\,q(x)\, p_g(x+y) \leq \sup_g [C_q C_{p_g}]^{1/2} \leq C_q  \]
for ignorance-respecting estimates, i.e.,
\begin{equation} \label{honest}
K_I \geq \left[ \int dx\,q(x)^2 \right]^{-1} .
\end{equation}
This is stronger than the lower bound (\ref{qmax}), and can be nontrivial even when $q_{\rm max}=\infty$.

\subsubsection{Continuous generators:}

Similar results hold for a shift generator $G$ with a continuous spectrum.  In particular, from  equation (\ref{herm}) of appendix A, 
identifying $G$ with $Y$,  a suitable scaling constant $K_I$ follows as  
\begin{equation} \label{kicont}
K_I^{-1} = \sup_{g,y} \langle g|\overline{M}_y|g\rangle = \sup_{g,y} \int dx\,q(x)\,\langle g|M_{x+y}|g\rangle .
\end{equation}
Here $|g\rangle$ ranges over all (typically degenerate) unit eigenkets appearing in any spectral decomposition of ${ \Gamma_g}$ (i.e., $|g\rangle \equiv|g,d\rangle$ for some $d$ in an orthogonal expansion ${ \Gamma_g} = \sum_d |g,d\rangle\langle g,d|$, where $d$ is an arbitrary degeneracy index).

It has not been possible at this time to prove a  general  relative resolution bound for continuous generators, analogous to (\ref{gendis}).  In particular, for continuous generators the ket $|g\rangle$ is not normalisable, so that $\langle g|M_{x+y}|g\rangle$ in (\ref{kicont}) does not correspond to some measurement probability density $p_g(x)$.  However, it is conjectured  that
\begin{equation} \label{con}
 \frac{\delta\hat{X}}{L} \geq \frac{\sqrt{\pi/2e}}{ 2L\langle |G-g|\rangle} ~~~~~~~~~~~~~{\rm (conjecture)}
\end{equation}
for such generators, if the prior distribution is uniform over an interval of length $L$.  Here, as always, the factor of 2 in the denominator can be dropped for a bounded spectrum for the choice $g=g_{\min}$.
The support for this conjecture arises from a correspondence between the cases of no prior information and covariant estimates, 
 as will now be detailed. 

First,  if the POVM $\{M_{\hat{x}}\}$ is covariant, then from equation (\ref{my})  it follows  that $\overline{M}_y=\int dx\,q(x)\,M_y = M_y$, for any prior distribution $q(x)$.  Hence, covariant estimates cannot make use of any prior information.  Conversely, any estimate that does makes use of prior information must be noncovariant.  

Second, for continuous generators,  $H(G)+H({M})\geq \ln \pi e$ for any covariant POVM ${ \cal M}$ from equation (\ref{gm}), where the bound is saturated when ${M}$ is canonically conjugate to $G$.  Hence one may take $K_I=\pi e$ for covariant estimates.  Moreover, for $m$ probe states, any covariant estimate of $X$ satisfies the same entropic uncertainty relation with $G$ replaced by $G_{ \rm T}=G_1+\dots G_m$.  Hence, using (\ref{gen}), (\ref{ki}), and $|G_1+\dots+G_m-mg| \leq |G_1-g|+\dots+|G_m-g|$, one has the rigorous bound
 \begin{equation} \label{cov}
   \delta\hat{X}_{\rm cov} \geq \frac{\sqrt{\pi/(2e)}}{2m\langle|  G  -g|\rangle} ,
\end{equation}
for covariant estimates,  where the expectation value is with respect to $\varrho_0$.  
As usual,  the denominator can be replaced by $m\langle  G -g_{ \rm min}\rangle$ if $G$ is bounded below. This bound includes covariant estimates based on entangling joint measurements; a class of covariant estimates based on independent measurements is also briefly discussed in appendix A.

Finally, since inequality (\ref{cov}) corresponds to no use of prior information, it can be interpreted in a limiting sense as a resolution bound relative to a prior distribution $q(x)$ which is uniform over the whole real line.  This corresponds to $L\rightarrow\infty$ in (\ref{con}), where more generally  the conjecture claims the same relative bound holds for prior distributions uniform over any finite interval of length $L$.

\section{Examples: discrete shift generators}

\subsection{Phase shift estimation}

For the case where the spectrum of the generator $G$ is a subset of the integers  one has $e^{iG\phi}=e^{iG(\phi+2\pi)}$, and hence the corresponding shift parameter $\Phi$ may be treated as a phase parameter, taking values on the unit circle.   It follows from equations (\ref{gen}), (\ref{ki}) and (\ref{qmax}) that 
\begin{equation} \label{phasegen}
\delta\hat{\Phi}  \geq \frac{(2\pi e^3)^{-1/2}\,K_I}{\langle 2|G-g|+1\rangle}\geq  \frac{(2\pi/ e^3)^{1/2}}{2\pi q_{\rm max}\,\langle 2|G-g|+1\rangle}\, ,
\end{equation}
where $q_{\max}\geq 1/2\pi$ is the maximum value of the prior probability density $q(\phi)$ for $\Phi$.  Moreover, if it is  known  {\it a priori} that the phase shift is confined to an interval of length $L$, then $q(x)=L^{-1}$ over the interval and vanishes elswhere, and hence from equation (\ref{gendis})  for $m$ identical systems  the relative resolution is bounded by
\begin{equation}\label{phasem}
\frac{\delta\hat{\Phi}}{L} \geq \frac{(2\pi e^3)^{-1/2}}{1+ 2m\langle \,|G_1-g|\, \rangle } .
\end{equation}

 As previously, the term $2|G-g|$ in the denominators of (\ref{phasegen}) and (\ref{phasem}) may be replaced by $G-g_{\rm min}$ when the spectrum is bounded below.  Hence for a single mode field, with number operator $N\geq 0$, one has $\delta\hat{\Phi}\geq (2\pi /e^3)^{1/2}/\langle N+1\rangle$ in the case of no prior information, with $q(\phi)= 1/2\pi = q_{\rm max}$.  However, strong numerical evidence has been given that the numerator $(2\pi /e^3)^{1/2}\approx 0.559$ for this case can be replaced by a best possible value of $\approx 1.376$, which is asymptotically achievable for large $\langle N\rangle$ on a suitable probe state via the canonical phase estimate $M_\phi = (2\pi)^{-1} \sum_{m,n}e^{-i(m-n)\phi}|m\rangle\langle n|$ \cite{prl,prep}.  Hence it is conjectured that the bounds (\ref{phasegen}) and (\ref{phasem}) are not tight, and that a similar replacement can be made in the case of arbitrary prior information. 

To exploit any prior information,  corresponding to a scaling constant $K_I<2\pi$  in (\ref{phasegen}),  is nontrivial.  Note that a covariant estimate is not suitable, as for any such estimate one has $\langle g|M_\phi|g\rangle= (2\pi)^{-1}$ for all eigenstates $|g\rangle$ of $G$ \cite{holevo,jmodopt}, yielding $K_I=2\pi$ from equation (\ref{kinverse}).  Indeed, equation (\ref{kinverse}) implies that a necessary condition for exploiting prior information is that {\it the estimate must return a nonuniform distribution over $[0,2\pi]$ when some number eigenstate is input as a probe state}.  This is counterintuitive, since such eigenstates are invariant under phase shifts and hence cannot generate any useful phase information.  However, it must be kept in mind that it is the actual probe state, $\rho_0$, rather than a notional probe state, that is relevant for actually estimating the phase.

For example, in the recently proposed phase estimation scheme of Rivas and Luis, applicable to small phase shifts of a single mode field generated by the photon number operator $N$, the estimate is proportional to the result of a homodyne measurement on the probe state \cite{rivas}.  Hence, if a number eigenstate $|n\rangle$ was to be input as a probe state, such a homodyne measurement would generate  the (clearly nonuniform)  statistics proportional to the quadrature distribution  $|\langle x|n\rangle|^2$.    Thus, the Rivas and Luis scheme satisfies the above necessary condition for exploiting prior information.  However, as noted by Rivas and Luis following their equation (29), while their estimate has an arbitrarily low   root mean square error $\delta\hat{\Phi}_0$ for a fixed phase shift value of zero, it can only  further achieve an error  $\delta \hat{\Phi}_{\phi} \approx \delta \hat{\Phi}_{0} \ll L$,  for each $\phi$ in an interval of length $L$ about  $\phi=0$ (corresponding to $\Delta \phi \ll \delta \phi$ in the notation of \cite{rivas}), if $\delta\hat{\Phi}_0\gg 1/N_{\rm  T}$, where  $N_{\rm  T}\equiv m \,{\rm tr}[N\varrho_0]$.  Thus, unfortunately, noting that the averaged root mean square error $\delta \hat{\Phi}$ over the interval is  $\approx \delta\hat{\Phi}_0$ in this case,  this scheme does not approach the lower bound in equation (\ref{phasem}) above,  nor even the numerically optimal bound $\delta \hat{\Phi}\geq 1.376/\langle N+1\rangle$ when prior information is {\it not} exploited \cite{prl}. 

The above  derivation establishes ultimate bounds  on phase resolution.  In particular, for any phase estimation scheme it is impossible to have  a scaling better than $1/q_{\rm max}$, or an asymptotic scaling better than  $1/\langle m|G|\rangle$.    However, it remains a challenge for further work to determine how closely the above phase estimation bounds can be be approached (up to some numerical factor), via a suitable measurement and probe state. 

Finally, it is worth noting that while $\langle G\rangle$ is a transparent measure of resources when $G$ is a photon number operator or similar, more generally the quantity $\langle|G-g|\rangle$ may not be.  However, it is straightforward to generalise the method used in section 2 to obtain bounds for resolution in terms of other quantities.  
For example, in the context of phase measurements, consider the shift generator  $G=\frac{1}{2}(N_A-N_B)$, where $N_A$ and $N_B$  are the number operators for respective single-mode fields input to a Mach-Zehnder interferometer \cite{mz}.  If the total number of input photons is bounded by some fixed maximum value, i.e.,   $N_A+N_B\leq N_{\rm max}$,  
then $G$ can only take $2N_{\rm max}+1$ distinct values: $0, \pm 1/2, \pm 1, \dots, \pm N_{\rm max}/2$. Hence $H(G)\leq \ln (2N_{\rm max}+1)$, and equations (\ref{xbound}) and (\ref{qmax})  yield the bounds
\begin{equation} \label{int}
\delta\hat{\Theta} \geq \frac{(2 \pi e)^{-1/2}K_I}{2N_{\rm max}+1},~~~~~~\frac{\delta\hat{\Theta}}{L} \geq \frac{(2 \pi e)^{-1/2}}{ 2m N_{\rm max} +1},
\end{equation}
 for the resolution of the corresponding shift parameter $\Theta\in[0,4\pi]$, 
  where $m$ and $L$ have the meanings introduced earlier.

\subsection{Time estimation for discrete Hamiltonians}

Time estimates correspond to the case $G=E/\hbar$, where $E$ is a Hamiltonian operator with lowest eigenvalue $\epsilon_0$.  Thus $G$ generates the time shift operator $e^{-iEt/\hbar}$.  For the case of a discrete spectrum, the Heisenberg bound (\ref{gen})  (recalling the factor of 2 can be removed for $g=\epsilon_0$), in combination with relations (\ref{ki}) and (\ref{qmax}), yields
\begin{equation} \label{time}
\delta\hat{T} \geq    \frac{ k_I}{1+ \langle E-\epsilon_0\rangle/\Delta_E} \geq \frac{1}{q_{\max}}\, \frac{(2\pi e^3)^{-1/2}}{1+ \langle E-\epsilon_0\rangle/\Delta_E} .
\end{equation}
Here, $\Delta_E$ denotes the smallest energy gap between distinct eigenvalues of $E$ (thus, the bound is only useful for $\Delta_E>0$), and $q_{\max}$ is the largest value of the prior density $q(t)$.  Further, if the time shift  is {\it a priori} uniformly distributed over an interval of length $\tau$  (which can be arbitrarily large if the system is not periodic), equation (\ref{gendis}) yields the scaling bound
\begin{equation} \label{timem}
\frac{\delta\hat{T}}{\tau} \geq \frac{(2\pi e^3)^{-1/2}}{1+ m\langle E-\epsilon_0\rangle/\Delta_E}\approx \frac{0.089}{1+ m\langle E-\epsilon_0\rangle/\Delta_E} .
\end{equation}
  where $m$ is the number of identically prepared copies. 
 
When the energy differences $\epsilon_j-\epsilon_k$ are incommensurate, the system will be almost periodic and $q(t)$ must be defined over the whole real line.  However, despite the nonexistence of a uniform prior distribution in this case, one can still define covariant time estimates  and show, for example, that one cannot typically extract more than 1 bit of information from such an estimate \cite{time}.  It is therefore expected that a noncovariant estimate is required to exploit any prior information.

An alternative resource of interest for bounding time resolution, particularly if the spectral structure is complex, is the number of energy eigenstates accessible to the probe state.  For example, if the  probe state is a $D$-level system, then the entropy of its energy distribution must satisfy $H(E)\leq \ln D$, implying via the entropic bound (\ref{xbound}) that $\delta\hat{T} \geq (2\pi e)^{-1/2}K_I/D$.  Hence, via (\ref{qmax}), one has the relative resolution bound 
\begin{equation} \label{d} 
\frac{\delta\hat{T}}{\tau} \geq    \frac{ (2\pi e)^{-1/2}}{mD} \approx \frac{0.242}{mD} 
\end{equation}
for a time shift uniformly distributed over $[0,\tau]$.

Note, however, that it is reasonable to expect that the actual time resolution of a discrete system will have a strong dependence on the detailed structure of the energy spectrum.  Hence,  the  above bounds  may  be well below what is  actually achievable. 

Finally, note that for a prior distribution uniform over an interval of length $\tau$, Tsang has very recently given the lower bound \cite{tsang}
\begin{equation} \label{qzz}
\frac{\delta\hat{T}}{\tau} \geq \frac{0.154\hbar}{\tau\langle E-\epsilon_0\rangle}\sqrt{1-\frac{0.329\hbar}{\tau\langle E-\epsilon_0\rangle}}
\end{equation}
for the relative resolution, under the constraint that $\tau\langle E-\epsilon_0\rangle\geq 0.690\hbar$.   For $\tau\Delta_E \leq 0.154(2\pi e^3)^{1/2}\hbar\approx 1.73\hbar$ this is asymptotically stronger, as a function of $\langle E-\epsilon\rangle^{-1}$,  than the relative resolution bound (\ref{timem}) (with $m=1$), and is weaker otherwise. Hence it provides an improved relative resolution bound for the case of  a sufficiently small energy gap, or a sufficiently small interval $\tau$ still satisfying the constraint.

\section{Examples: continuous shift generators}

\subsection{Time estimation for continuous Hamiltonians}

For Hamiltonians having a continuous spectrum, equation (\ref{gen}) (recalling the factor of 2 can be dropped for $g=\epsilon_0$), together with equation (\ref{ki}), yields
\begin{equation} \label{timecon}
\delta\hat{T} \geq    \frac{(2\pi e^3)^{-1/2}K_I}{ \langle E\rangle -\epsilon_0} ,
\end{equation}
where $K_I$ is given in (\ref{kicont}).  Further, for any covariant time estimate one has
\begin{equation} \label{tcov}
\delta\hat{T}_{\rm cov} \geq \frac{\sqrt{\pi/({2e})}}{m}\,\frac{\hbar}{\langle E\rangle-\epsilon_0} \approx \frac{0.76\,\hbar/m}{\langle E\rangle-\epsilon_0} 
\end{equation}
from (\ref{cov}).  Finally, if conjecture (\ref{con}) is correct, then  (again dropping the factor of 2) the relative resolution bound
\[ \frac{\delta\hat{T}}{\tau} \geq  \frac{\sqrt{\pi/2e}\,\hbar}{m\tau\langle E-\epsilon_0 \rangle}~~~~~~~~~~~~{\rm (conjecture)} \]
holds for a prior probability density $q(t)$ uniform over an interval of length $\tau$.
 
It would be of great interest to determine how closely the bound (\ref{tcov}) can be approached, via a canonical time measurement on a suitable probe state.  Note that the variance of the canonical time distribution does not exist if the energy distribution of the probe state has a nonzero groundstate component  \cite{holevo}.  Hence, such probe states would require a different measure of time resolution  ---  e.g., the ensemble length of the error distribution, $\exp[H(\hat{T}-T)]$, which may be bounded  from below  via inequalities (\ref{hbound}) and (\ref{hmax}).

\subsection{Spatial displacement estimation}

As a final example, consider the case of estimation of the displacement of a quantum system in some direction, corresponding to the generator $G=P/\hbar$ for the momentum in that direction.  The general resolution bound 
\begin{equation}
 \delta\hat{X} \geq \frac{\hbar}{2}\,\frac{(2\pi e^3)^{-1/2}K_I}{\langle  |P-p| \rangle}  
 \end{equation}
follows from (\ref{gen}) and (\ref{ki}), while for any covariant estimate one has
\begin{equation} \label{covx}
\delta\hat{X}_{\rm cov} \geq \frac{\hbar}{2}\,\frac{\sqrt{\pi/(2e)}}{ \langle  |P-p| \rangle} 
\end{equation}
from (\ref{cov}),  for the case of a single copy, $m=1$.   
The conjecture (\ref{con}) implies a similar bound for the case of a prior distribution uniform over any finite interval, suggesting that measurement of  the position observable $Q$ conjugate to $P$, with POVM elements  $M_q = |q\rangle\langle q|$,  is always optimal in this case. 

The above bounds are valid for all values of the reference momentum  $p$.  However, a variational calculation shows they are strongest when  $p$  is chosen to be the median value  $p_0$  of the momentum distribution $p(k)$, i.e, when  $\int_{-\infty}^{p_0} dk\, p(k) = 1/2$.  Note that the mean and median values are identical for the case of symmetric distributions. 

It is of interest to note  that the  covariant resolution bound (\ref{covx}) can be weaker or stronger than the  Cramer-Rao related  bound, $\hbar/(2\Delta P)$, following from (\ref{cr}) \cite{WisMil10,helstrom,holevo,cr}.  For example, for probe states with a Gaussian momentum distribution, $p(k)\sim \exp[-k^2/(2\sigma^2)]$, one finds $\Delta P=(\pi/2)^{1/2}\langle|P|\rangle$, implying that the   bound (\ref{cr})  is stronger by a factor of 1.05.  Conversely, for a probe state with an exponential momentum distribution, $p(k) \sim\exp[-\beta|k|]$, one finds $\Delta P=\sqrt{2}\langle|P|\rangle$, implying that the  bound (\ref{cr})   is weaker by a factor of 1.08.
\section{Including the effects of noise}

The presence of noise is expected to decrease the accuracy of any estimate, and hence to increase the lower bounds in the previous sections.
Consider, for example, the very simple case in which independent noise is added to the measurement outcome $\hat{X}$. Denoting the noise variable by $Z$, the entropy power inequality  \cite{power} and equation (\ref{hbound}) imply 
\[ e^{2H(\hat{X}+Z-X)} \geq e^{2H(\hat{X}-X)}+e^{2H(Z)} \geq K_I^2e^{-2H(G)}+e^{2H(Z)} .\]
Hence, as per the derivation of equation (\ref{xbound}), it follows that
\begin{equation}
\delta\hat{X}_{\rm  noisy} \geq \left[ (\delta\hat{X}_{\rm bound})^2 + (2\pi e)^{-1}e^{2H(Z)}  \right]^{1/2} ,
\end{equation} 
where $\hat{X}_{\rm  noisy}:=\hat{X}+Z$ and $\delta\hat{X}_{\rm bound}$ denotes any of the lower bounds of the previous sections.  Noise thus increases the minimum possible resolution.

A more physical approach is to consider processes that add noise directly to the probe state, and to use stronger entropic uncertainty relations which depend on the probe state.  For example, for a rank-1 {\it discrete} generator $G$, inequality (\ref{rank1}) of Appendix A effectively replaces $K_I$ by $K_Ie^{S[\rho_0]}$, where $S[\rho_0]$ is the von Neumann entropy of the probe state. Hence, replacing $\rho_0$ by the noisy probe state $\Pi(\rho_0)$ for some completely positive map $\Pi$, the generic lower bound (\ref{gen}) generalises to
\begin{equation} \label{gen1}
\delta \hat{X} \geq \frac{k_I}{1+2\Delta^{-1}\langle \,|G-g|\, \rangle_\Pi } e^{S[\Pi(\rho_0)]}
\end{equation}
for such generators, where $\langle \cdot\rangle_\Pi$ denotes an average with respect to $\Pi(\rho_0)$. Similarly, the relative resolution bound (\ref{gendis}) generalises to
\begin{equation}\label{gendis1}
\frac{\delta\hat{X}}{L} \geq \frac{(2\pi e^3)^{-1/2}}{1+ 2\Delta^{-1}\langle\,|G-g|\, \rangle_\Pi } e^{S[\Pi(\rho_0)]} .
\end{equation}
(one is limited to the case $m=1$, since $G_{ \rm T}=G_1+\dots+G_m$ is not rank-1 for $m>1$). For a rank-1 {\it continuous} generator $G$, the covariant bound (\ref{cov}) generalises to
 \begin{equation} \label{cov1}
   \delta\hat{X}_{\rm cov} \geq \frac{\sqrt{2\pi/e^3}}{2\langle|G-g|\rangle_\Pi}e^{S[\Pi(\rho_0)]} ,
\end{equation}
via uncertainty relation (\ref{rank3}) of Appendix A.
As always, the factor of 2 in the above denominators may be removed for the choice $g=g_{\rm min}$.

For example, let $N$ be the photon number operator of a single mode field subject to Gaussian noise, where the noise is described by the completely positive map \cite{vourdas}
\[  \Lambda(\rho)  
= \int d^2\alpha\, (\pi n_{ \lambda})^{-1} e^{-|\alpha|^2/n_{ \lambda}} D(\alpha)\rho D(\alpha)^\dagger ,\]
and $D(\alpha)=\exp(\alpha a^\dagger-\alpha^*a)$ denotes the Glauber displacement operator.  The parameter $n_{ \lambda}$ characterises the average number of photons added to the field, i.e., $\langle N\rangle_{ \Lambda} = \langle N\rangle +n_{ \lambda}$, while the entropy of the field is bounded, both for pure and mixed states, by \cite{gauss}
\[   S[ \Lambda(\rho)  ] \geq \ln (1+n_{ \lambda}) + n_{ \lambda} \ln (1+1/n_{ \lambda}) \geq \ln (1+n_{ \lambda}) . \]
 Combining this with the relative resolution bound (\ref{gendis1}) then yields a  `noisy' bound
\begin{equation}\label{phasen}
\frac{\delta\hat{\Phi}}{L} \geq \frac{(2\pi e^3)^{-1/2}}{1+ \langle N \rangle/(n_{ \lambda}+1) } 
\end{equation}
for the relative resolution of any phase estimate, for a prior distribution uniform over some interval of width $L$.  It is seen that the resolution becomes poor for sufficiently large noise.

\section{Discussion}

The results of the paper establish a rigorous, nonasymptotic and constraint-free  lower bound 
for parameter estimation which is in the form of the Heisenberg limit and which takes   prior information into account.  The fundamental bound (\ref{gen}) implies that asymptotic scaling better than $1/\langle |G|\rangle$ is impossible, while bound (\ref{gendis}) for discrete generators further demonstrates that,  for shifts randomly distributed over some interval,  asymptotic scaling better than $1/m$ is impossible for the relative resolution, where $m$ is the number of probe states (and entangling joint measurements are permitted).   Bound (\ref{cov}) for continuous generators implies a similar $1/m$ limit is unavoidable for the case of covariant estimates. It has also been shown how the effects of noise may be quantified in section 5, including resolution bound (\ref{phasen}) for phase estimates on a single mode field subjected to Gaussian noise.

Examples have been given for estimates of phase shifts, time shifts and spatial displacement in sections 3 and 4.  These sections also give examples of how the basic method of section 2 may be applied to obtain resolution bounds in terms of alternative resources, such as the total available photon number in equation (\ref{int}) and the energy support of the probe state in equation (\ref{d}).  For the case of discrete generators with a finite minimum eigenvalue, the corresponding relative resolution bound (\ref{timem}) may be stronger or weaker than the recent constrained bound (\ref{qzz}) due to Tsang (section 3.2). 

The fundamental tool used to obtain the above resolution bounds is equation (\ref{hbound}) for the entropy of the error in the estimate, $H(\hat{X}-X)$.  As noted briefly in section 2.2, the exponential of this entropy may be useful as an alternative possible measure of resolution \cite{vol}.  Further, this measure has tighter corresponding bounds, as it avoids the use of the relation between $\delta\hat{X}$ and $H(\hat{X}-X)$ (which is only saturated for Gaussian distributions), required for obtaining equation (\ref{xbound}). 

Another interesting measure of resolution to consider is the mutual information between the shift and its estimate, $H(\hat{X}:X)$ \cite{power}.  While mutual information is not dealt with directly in this paper, the relative resolution bounds (\ref{gendis}), (\ref{phasem}), (\ref{int}), (\ref{timem}), (\ref{d}), (\ref{gendis1}) and (\ref{phasen}) do allow an  approximate upper bound to be derived for  $H(\hat{X}:X)$, for discrete generators, whenever the prior distribution is uniform over some  sufficiently large  interval.  
In  particular, the number of distributions of width $\delta\hat{X}$ that can be distinguished without error, over an interval of width $L$, is approximately $L/\delta\hat{X}$.  The corresponding mutual information, i.e.,  the corresponding number of bits that can be encoded by the shift parameter $X$ and distinguished by the estimate $\hat{X}$ \cite{power}, is
therefore $H(\hat{X}:X)\approx \log_2 [L/\delta\hat{X}]$, i.e., the logarithm of the reciprocal of the relative resolution.  Thus the above mentioned relative resolution bounds place an approximate upper  bound  on the mutual information.  For example, for estimates of phase shifts uniform over an interval of length $L$, generated by the photon number of a single mode field subjected to Gaussian noise, one has the approximate upper bound
\begin{equation}
H(\hat{\Phi}:\Phi)\lesssim \log_2 \left[1+ \langle N \rangle/(n_{ \lambda}+1)\right] +\frac{1}{2} \log_2 [2\pi e^3]
\end{equation}
for mutual information from equation (\ref{phasen}).  If the conjectured bound (\ref{con}) is correct, one may similarly obtain estimates of mutual information for continuous generators.  

As noted above, the resolution can scale no better than inversely with the number of probe states, $m$, even when entangling joint measurements are permitted.  As noted in the introduction, this contrasts with the  $m^{-1/2}$ scaling of the Cramer-Rao related bound  (\ref{cr}).     
Note that for $m=1$, the bounds of this paper can be numerically weaker or stronger than (\ref{cr}) (section 4.2).  

It has been seen in sections 2-4 that covariant estimates do not exploit any prior information that may be available.  Hence it is only possible to approach the generic resolution bounds (\ref{gen}) and ({\ref{gendis}) via noncovariant estimates.  Further, as noted in section 3.1, a necessary condition for making use of prior information is that the measurement scheme must return a nonuniform distribution when some eigenstate of the generator is input as a probe state.  While the recently proposed scheme of Luis and Rivas meets this condition, it does not approach the corresponding bound (\ref{phasem}) (section 3.1).  It therefore remains an important challenge for future work to determine how closely the various lower bounds of this paper can be approached.  

It is also hoped that future work will settle the conjectures made regarding the relative resolution bound (\ref{con}) for continuous generators in section 2.3, the improvement in scaling factors for phase estimates in section 3.1, and the strong entropic uncertainty bound (\ref{rank2}) in Appendix A.

Finally, it is noted that the extensions of various entropic uncertainty relations to continuous observables, obtained in Appendix A, will find application beyond the realm of quantum metrology.

{\bf Acknowledgment}: This work was supported by the ARC Centre of Excellence CE110001027.

\appendix

\section{Entropic uncertainty relations involving continuous POVMs}

\subsection{One continuous observable}

On a finite-dimensional Hilbert space, the entropies of two observables $A$ and $B$, corresponding to finitely-valued POVMs $\{ A_j\}$ and $\{ B_k\}$, satisfy the entropic uncertainty relation \cite{para, entreview}
\begin{equation} \label{para}
H(A) + H(B) \geq - 2 \ln \max_{j,k} \left\| A_j^{1/2}B_k^{1/2} \right\|_\infty, 
\end{equation}
where $\left\|X\right\|_\infty$ denotes the largest singular value of $X$, i.e., the square root of the largest eigenvalue of $X^\dagger X$. 

To extend this relation to the case where one of the observables is continuously valued, first consider some observable $C$ taking continuous values in some {\it compact} set, with corresponding POVM $\{ C_\theta\}$, and partition the range of $\theta$ into a finite number of nonoverlapping bins $\{P_k\}$ of equal size $\epsilon$.  Define the discrete observable $C^P$ associated with the partition via the POVM $\{C^P_k\}$ with
$C^{P}_k := \int_{P_k} d\theta \,C_\theta$.  Then, for any probability density $p(\theta)$ of $C$, there is a corresponding well-defined discrete probability distribution $p_k:=\int_{P_k} d\theta \, p(\theta)$ (equal to the probablity of $\theta\in P_k$), and an associated piecewise-continuous probability density $\tilde{p}(\theta)$ given by replacing $p(\theta)$ by its average value over the bin $P_k$ for $\theta\in P_k$, i.e., 
\[ \tilde{p}(\theta) := \epsilon^{-1} \int_{P_k} d\theta \, p(\theta) = \epsilon^{-1} p_k~~{\rm for~~}\theta\in P_k. \]
Note that the probability of $\theta\in P_k$ is identical for both $p(\theta)$ and $\tilde{p}(\theta)$, implying the latter converges in distribution to the former in the limit $\epsilon\rightarrow 0$. Note also that
the entropy of $C^{P}$,
$H(C^{P}) = -\sum_k p_k \ln p_k$,
can be rewritten using $\sum_k p_k=1$ as 
\[ H(C^{P}) = -\ln\epsilon   - \int d\theta\, \tilde{p}(\theta) \ln \tilde{p}(\theta) . \]

The entropic uncertainty relation (\ref{para}) for observables $A$ and $C^P$ gives
\begin{eqnarray*} 
H(A) + H(C^{P}) &\geq& -\max_{j,k,\psi} \ln\langle\psi | A_j^{1/2} C^{P}_k A_j^{1/2}|\psi\rangle\\
&=& - \max_{j,k,\psi} \ln\left[ \langle\psi_j |  C^{P}_k |\psi_j\rangle \,\langle\psi|A_j|\psi\rangle\right]\\
&=& -\max_{j,\theta,\psi} \ln \left[ \epsilon \,\tilde{p}_{\psi_j}(\theta)\,\langle\psi|A_j|\psi\rangle\right]\\
&=& -\ln \epsilon - \max_{j,\theta,\psi} \ln \left[  \tilde{p}_{\psi_j}(\theta)\,\langle\psi|A_j|\psi\rangle\right],
\end{eqnarray*}
where $|\psi_j\rangle:=A_j^{1/2}|\psi\rangle / \langle\psi|A_j|\psi\rangle^{1/2}$, and $\tilde{p}_{\psi_j}(\theta)$ is defined analogously to $\tilde{p}(\theta)$ above, with respect to the probability density $p_{\psi_j}(\theta)=\langle\psi_j|C_\theta|\psi_j\rangle$ (and thus converges in distribution to $p_{\psi_j}(\theta)$ in the limit $\epsilon\rightarrow 0$).   Using the above expression for $H(C^{P})$ then yields, taking the limit $\epsilon\rightarrow 0$,  
\begin{eqnarray} \nonumber 
H(A) + H(C) &\geq& -\sup_{j,\theta,\psi} \ln  \left[\langle\psi_j|C_\theta|\psi_j\rangle\, \langle\psi|A_j|\psi\rangle\right]  \\
\label{eurac} 
&=& -\sup_{j,\theta,\psi} \ln \langle\psi| A_j^{1/2}C_\theta A_j^{1/2}|\psi\rangle\\  
\label{euracfin}
&=& -2 \sup_{j,\theta} \ln \| A_j^{1/2} C_\theta^{1/2} \|_\infty ,
\end{eqnarray}
whenever $H(C)$ exists, thus generalising (\ref{para}).  

Uncertainty relations (\ref{eurac}) and (\ref{euracfin}) may be further extended to the case of infinite-dimensional Hilbert spaces, whenever the left hand side exists, by considering the limit of a series of projections of the observables onto finite Hilbert spaces.  They similarly extend to the case of a countably infinite POVM $\{A_j\}$, whenever $H(A)$ exists, by considering the limit of the sequence of finite POVMs $\{ A_1,A_2,\dots,A_d,1-\sum_{j=1}^d A_j\}$ as $d\rightarrow\infty$.  Finally, they also extend to the case of a non-compact range of $C$, whenever the left hand side exists, by representing the range as the limit of a series of compact sets $S$, and replacing $\{ C_\theta\}$ by a corresponding series of POVMs $\{C_\theta, \theta\in S\}\cup \{1-C_S\}$, where $C_S:=\int_S d\theta\,C_\theta$. 
Thus, (\ref{eurac}) is valid for any discrete-valued observable $A$ and continuously-valued observable $C$, whenever $H(A)+H(C)$ exists,  and similarly for (\ref{euracfin}) if $C_\theta^{1/2}$ is well defined.

For example, for an angular momentum component $J_z$ and its corresponding conjugate angle $\Phi$, inequality in (\ref{eurac}) reduces to the known result
\[ H(J_z) + H(\Phi) \geq -\sup_{j,\phi,\psi} \ln \left[ |\langle\psi|j\rangle|^2|\langle j|\phi\rangle^2\right] =\ln 2\pi\hbar , \]
first given by Bialynicki-Birula and Mycielski \cite{bbm}, which is saturated by eigenstates of $J_z$.

\subsection{Two continuous observables}

Uncertainty relation (\ref{euracfin}) provides the basis for extending to the case of two continuous-valued observables  $X$ and $Y$ corresponding to POVMs $\{X_x\}$ and $\{Y_y\}$ respectively.   The procedure is similar to the foregoing.  In particular, partitioning the range of $x$ into bins $P_j$ of equal size $\epsilon$, one has the corresponding discrete POVM ${\cal  A}\equiv\{A_j\}$ with $A_j:=\int_{P_j}dx\,X_j$. The entropy of  the corresponding observable $A$,  given  a continuous probability density $p(x)$ of $X$, is then that of the discrete distribution $p_j:=\int_{P_j} dx\,p(x) = \epsilon \,\tilde{p}(x)$, where the second expression defines the piecewise continuous probability density $\tilde{p}(x)$.  Substitution into (\ref{euracfin}), with $C\equiv Y$ and assuming $Y_y^{1/2}$ is well-defined, gives
\begin{eqnarray*}
-\ln\epsilon - \int dx\,\tilde{p}(x) \ln \tilde{p}(x)+H(Y) &\geq& -2 \sup_{j,y} \ln \| A_j^{1/2}Y_y^{1/2}\|_\infty\\
&=& -\sup_{j,y,\psi} \ln \,\langle\psi|Y_y^{1/2}A_jY_y^{1/2}|\psi\rangle\\
&=& -\sup_{j,y,\psi} \ln \left[  \epsilon \,\tilde{p}_{\psi_y}(x)\,\langle\psi|Y_y|\psi\rangle\right],
\end{eqnarray*}
where $ |\psi_y\rangle :=Y_y^{1/2}|\psi\rangle/\langle\psi|Y_y|\psi\rangle^{1/2}$ and $\tilde{p}_{\psi_y}(x)$ is defined analogously to $\tilde{p}(x)$ above, with respect to the probability density $p_{\psi_y}(x)=\langle\psi_y|X_x|\psi_y\rangle$.
Taking the limit $\epsilon \rightarrow 0$ then gives
\[
H(X) + H(Y) \geq -\sup_{x,y,\psi} \ln \left[  \,\langle\psi_y| X_x|\psi_y\rangle\,\langle\psi|Y_y|\psi\rangle\right] 
\]
\begin{equation} \label{eurxy}
~~~~~~~~~= -\sup_{x,y,\psi} \ln \langle \psi|Y_y^{1/2}X_xY_y^{1/2}|\psi\rangle = -2  \sup_{x,y} \ln \| X_x^{1/2}Y_y^{1/2}\|_\infty 
\end{equation}
whenever the entropies and the relevant square roots are well defined.  

Indeed, this uncertainty relation can also be applied in some instances when the square roots are not well defined, via taking appropriate limits.
For example, for conjugate position and momentum observables $Q$ and $P$, with eigenkets $|q\rangle$ and $|p\rangle$ respectively, and any $\epsilon >0$, define
the `averaged' momentum observable $\tilde{P}$ with POVM $\{\tilde{P}_p\}$ via $\tilde{P}_p:=(2\epsilon)^{-1}\int_{-\epsilon}^\epsilon dk\,|p+k\rangle\langle p+k|$.  Then, $\tilde{P}_p^{1/2}=(2\epsilon)^{1/2}\tilde{P}_p$ from the spectral theorem, yielding
\begin{eqnarray*}
 \langle\psi|\tilde{P}_p^{1/2}Q_q\tilde{P}_p^{1/2}|\psi\rangle &=&|\langle q|\tilde{P}_p^{1/2}|\psi\rangle|^2
= \frac{1}{2\epsilon} \left| \int_{-\epsilon}^\epsilon dk\,\langle q|p+k\rangle \langle p+k|\psi\rangle \right|^2\\
&\leq& \frac{1}{2\epsilon}  \int_{-\epsilon}^\epsilon dk\,|\langle q|p+k\rangle|^2 \int_{-\epsilon}^\epsilon dk\,|\langle p+k|\psi\rangle|^2 ,
\end{eqnarray*}
where the last line follows from the Schwarz inequality.  The first integral evaluates to $2\epsilon/(2\pi\hbar)$, while the second is never greater than unity for any normalised state $\psi$. Hence, substituting $X=Q$ and $Y=\tilde{P}$ into the first inequality of uncertainty relation (\ref{eurxy}), and taking the limit $\epsilon\rightarrow 0$, yields 
\begin{equation} \label{eurxp}
H(Q) + H(P) \geq\ln 2\pi\hbar .  
\end{equation}
Note that the lower bound is not optimal, although it is close to the optimal bound 
$\ln \pi e\hbar$,
saturated by Gaussian pure states \cite{entreview,bbm}.  However, the same lower bound is optimal for the related tight uncertainty relation (\ref{rank3}) below.  

 More generally, if the POM $\{Y_y\}$ is a continuous projection-valued measure corresponding to some Hermitian operator, then although $Y_y^{1/2}$ is not well defined, an entropic uncertainty relation may be obtained via a similar limiting approach. In particular, in such a case $Y_yY_{y'}=\delta(y-y')Y_y$,
implying that $\tilde{Y}_y:=(2\epsilon)^{-1}\int_{-\epsilon}^{\epsilon} dz\,Y_{y+z}$ satisfies $\tilde{Y}^{1/2}=(2\epsilon)^{1/2}\tilde{Y}_y$.  Applying the first inequality in ({\ref{eurxy}) to $X$ and $\tilde{Y}$ yields
\[ H(X) + H(\tilde{Y}) \geq -\sup_{x,y,\psi} \ln \left[  \,\langle\tilde{\psi}_y| X_x|\tilde{\psi_y}\rangle\,\langle\psi|\tilde{Y}_y|\psi\rangle\right], \]
with $|\tilde{\psi}_y\rangle :=\tilde{Y}_y^{1/2}|\psi\rangle/\langle\psi|\tilde{Y}_y|\psi\rangle^{1/2}= (2\epsilon)^{1/2}\tilde{Y}_y|\psi\rangle/\langle\psi|\tilde{Y}_y|\psi\rangle^{1/2}$.  Noting
\[ \langle\psi|\tilde{Y}_y|\psi\rangle = (2\epsilon)^{-1}\int_{-\epsilon}^{\epsilon} \langle\psi|{Y}_y|\psi\rangle \leq (2\epsilon)^{-1} , \]
the limit $\epsilon\rightarrow 0$  gives $H(X) + H(Y) \geq -\sup_{x,y,\psi} \ln \langle\psi| Y_yX_xY_y|\psi\rangle/\langle\psi| Y_y|\psi\rangle$.
Finally, $Y_y|\psi\rangle$ is always proportional to some unit eigenket of $Y$, i.e., $Y_y|\psi\rangle = \alpha|y,d\rangle$, where $d$ is a degeneracy index in some orthogonal expansion $Y_y=\sum_d |y,d\rangle\langle y,d|$ of $Y_y$ (noting such expansions are invariant under unitary transformations of the degeneracy basis), and the uncertainty relation
\begin{equation} \label{herm}
H(X) + H(Y) \geq -\sup_{x,y} \ln \, \langle y|X_x|y\rangle
\end{equation}
 immediately follows, where $|y\rangle$ ranges over all unit eigenkets of $Y_y$.  This generalisation of (\ref{eurxp}), holding for any projection-valued measure $\{Y_y\}$, is of particular relevance to generators with continuous spectra (section 2.3.2).


\subsection{One rank-1 observable}

When the observable $A$ in relation (\ref{para}) is rank 1, i.e., when $A_j=|a_j\rangle\langle a_j|$ for some set of (not necessarily  normalised) kets $\{|a_j\rangle\}$, then one has the stronger uncertainty relation \cite{coles}
\begin{equation}  \label{coles}
H(A) + H(B) \geq - \ln \max_{j,k} \left\| A_j^{1/2}B_k^{1/2} \right\|_\infty + S[\rho],
\end{equation}
where $S[\rho]$ denotes the von Neumann entropy $-{\rm tr}[\rho\ln\rho]$ of the density operator $\rho$ generating the statistics of $A$ and $B$.  This recent result, by Coles et al. \cite{coles}, generalises an earlier version by Berta et al. for the case of two rank-1 projection valued observables $A$ and $B$ \cite{berta}.

Using the same methods as in appendix A.1, this relation can be similarly generalised to infinite Hilbert spaces and one continuous observable, to give
\begin{equation} \label{rank1}
H(A) + H(C) \geq  -\sup_{j,\theta,\psi} \ln \langle \psi|A_j^{1/2}C_\theta A_j^{1/2}|\psi\rangle  +S[\rho]
\end{equation}
whenever the left hand side exists, for any discrete-valued rank-1 observable $A$ and any continuously-valued observable $C$.  

Unfortunately, one cannot analogously generalise (\ref{eurxy}) via the methods of appendix A.2, as these methods rely on use of an observable $A$ which is not rank 1.  However, it is conjectured here that such a generalisation exists, with
\begin{equation} \label{rank2}
H(X) + H(Y) \geq  -\sup_{x,y,\psi} \ln \langle \psi|Y_y^{1/2}X_xY_y^{1/2}|\psi\rangle + S[\rho]
\end{equation}
for any two continuously valued observables $X$ and $Y$, providing $X$ is rank 1, both entropies exist, and $Y_y^{1/2}$ is well defined.  

The above conjecture can be proved for the special case of conjugate position and momentum observables, using an approach of Pegg \etal in which $Q$ and $P$ are represented by approximating them as discrete rank-1 observables on a $D$-dimensional Hilbert space and taking the limit $D\rightarrow \infty$ \cite{pegg}. In particular, substituting the discrete observables into (\ref{coles}) and taking this limit yields
\begin{equation} \label{rank3}
H(Q) + H(P) - S[\rho] \geq \ln 2\pi\hbar ,
\end{equation}
whenever the left hand side exists.  The same method generalises to the case of conjugate $n$-vectors ${\bf Q}$ and ${\bf P}$, with the right hand side of the above relation being multiplied by $n$. 
This result proves the conjecture in equation (47) of \cite{vol}, which was made on the basis of a semiclassical argument.  Note that, in contrast to inequality (\ref{eurxp}), the bound in inequality (\ref{rank3}) is tight, being saturated in the limit of equilibrium states in the high temperature limit \cite{vol}.

\subsection{Covariant observables}

The entropic uncertainty relation
\begin{equation} \label{gm}
H(G) +H(M) \geq \ln \pi e , \end{equation}
will be obtained here, for any observable $M$ covariant with respect to a continuous Hermitian observable $G$, and also a generalisation to estimates based on $m$ repeated measurements, as required for equation (\ref{cov}) of the text.

First, let $F$ denote the observable canonically conjugate to $G$, with POVM $\{F_x\}$ given by \cite{holevo,time}
\[ F_x := e^{-iGx}F_0e^{iGx},~~~F_0:= (2\pi)^{-1} \sum_d \int dg\, dg'\, |g,d\rangle\langle g',d| ,\]
where the projection ${ \Gamma_g}$ has the orthogonal expansion ${ \Gamma_g}=\sum_d |g,d\rangle\langle g,d|$.  Now, any density operator $\rho$ may be formally mapped to a density operator $\rho^*$ of a 1-dimensional particle, with position coordinate $Q$ and eigenkets $|q\rangle$, via
$\rho^*:= \sum_d \int dq\,dq' |q\rangle\langle q'|\,\langle q,d|\rho|q',d\rangle$,
(where one takes $|g,d\rangle:=0$ for $g$ outside the spectrum of $G$).  By construction, the probability distributions of $F$ and $G$ for $\rho$ are identical to the probability distributions of $Q$ and $P$ for $\rho^*$, where $P$ is the momentum observable conjugate to $Q$ in units such that $\hbar=1$.  From the known entropic uncertainty relation for $Q$ and $P$ \cite{entreview,bbm}, it immediately  follows that
\[ H(F)+H(G) = H(Q) + H(P) \geq \ln \pi e . \]
Further, the probability distribution of any covariant observable $M$ for some state $\rho$ is equal to the probability distribution of the conjugate observable $F$ for some corresponding state $\rho'$, where $\rho$ and $\rho'$ have the same probability distribution for the observable $G$ \cite{time}. Hence the above bound also holds with $F$ replaced by $M$, yielding (\ref{gm}) as desired.

It is of interest to consider a class of covariant estimates $m$ based on independent measurements of the covariant observable $M$, made on $m$ respective copies of the probe state. Let $\hat{x}=f(\hat{x}_1,\hat{x}_2,\dots,\hat{x}_m)$ denote the corresponding estimate of the shift parameter, where $\hat{x}_j$ denotes the individual estimate given by  the $j$-th measurement.  It will be assumed that the estimate is shift-invariant, i.e., that the function  $f$ satisfies the identity 
\begin{equation} \label{fid} f({x}_1+y,{x}_2+y,\dots,{x}_m+y)=f({x}_1,{x}_2,\dots,{x}_m)+y. 
\end{equation} 
This is satisfied, for example, by any weighted mean $f=w_1x_1+\dots w_mx_m$ with $w_1+\dots+w_m=1$. Differentiating with respect to $y$ and taking $y=0$ gives the equivalent condition $\sum_j (\partial f/\partial x_j)=1$.

To show that such an estimate is covariant with respect to the corresponding shift generator $G_{ \rm T}=G_1+\dots+G_m$, let ${\bf x}$ denote the vector $(x_1,\dots,x_m)$, and define $S_x:=\{ {\bf x}: f({\bf x})=x\}$.  The POVM $\{ M^{J}_{\hat{x}}\}$ corresponding to the joint estimate is then given by
$M^{J}_{\hat{x}} = \int_{S_{\hat{x}}} d{\bf x} \,M_{\hat{x}_1}\otimes \dots \otimes M_{\hat{x}_m}$ and therefore
\begin{eqnarray*}
 e^{-iG_{ \rm T}x}M^{J}_ye^{iG_{ \rm T}x} &=& \int_{S_{y}} d{\bf y} \,e^{-iG_1x}M_{{y}_1}e^{iG_1x}\otimes \dots \otimes e^{-iG_mx}M_{{y}_m}e^{iG_mx}\\
&=& \int_{S_{y}} d{\bf y} \,M_{{x+y}_1}\otimes \dots \otimes M_{x+{y}_m}\\
&=& \int_{S_{x+y}} d{\bf y} \,M_{{y}_1}\otimes \dots \otimes M_{{y}_m} = M^J_{x+y} ,
\end{eqnarray*}
where the last line follows via the change of variables $y_m\rightarrow y_m-x$ and using the above identity for $f$.  Hence $M^J$ is covariant with respect to $G_{ \rm T}$.

\section{Entropy bounds for fixed $\langle|G-g|\rangle$}

\subsection{Discrete generators}

First, consider the special case $G$ where the spectrum of $G$ is some subset of the integers.  Now, the maximum entropy of any distribution over the integers for a fixed value of $\langle |n-n_0|\rangle$ corresponds to maximising the variational quantity
\[ J = -\sum_n p_n\ln p_n +\alpha\sum_np_n +\beta\sum_n |n-n_0| p_n ,\]
where $\alpha$ and $\beta$ are Lagrange multipliers.  
It is convenient to work with the displaced distribution $q_n:=p_{n+n_0}$, for which the variational equation $\delta J/\delta q_n=0$ has the solution $q_n = Av^{|n|}$, for suitable positive constants $A$ and $v$ determined by the contraints.  One easily finds that $A=(1-v)/(1+v)$ and $\overline{n}:=\langle |n-n_0|\rangle = \sum_n |n|q_n = 2v/(1-v^2)$. Further, 
$H_{\rm max}(G) = -\sum_n q_n [\ln A + |n|\ln v] = -\ln A - \overline{n}\ln v$.  Inverting  the relation between $\overline{n}$ and $v$ gives $v=(1+1/\overline{n}^2)^{1/2} - 1/\overline{n}=e^{-x}$, where $\sinh x=1/\overline{n}$, and the maximum entropy simplifies to
\begin{equation} \label{number}
H_{\rm max}(G) = \ln\left[(\sqrt{\overline{n}^2+1}+\overline{n}\right] + x/\sinh x \leq \ln [2\overline{n}+1] + 1, 
\end{equation}
where the final inequality follows by adding a term $2\overline{n}$ under the square root and noting that $x\leq \sinh x$ (e.g., from the Taylor series expansion of $\sinh x$).  This corresponds to the bound (\ref{hmax}) with $\Delta =1=c$.  Further, if the spectrum of $G$ is bounded below by $n_{\rm min}$, a similar calculation for the choice $n_0=n_{\rm min}$ leads to $H_{\rm max}(G)\leq \ln (\overline{n}+1) +1$ \cite{prl}, i.e., to replacement of the factor of 2 in (\ref{number}) by unity.

More generally, let $G$ have an arbitrary discrete spectrum, and let $\{ g_n\}$ denote the distinct eigenvalues of $G$ in increasing order. Thus, the minimum spectral gap is given by $\Delta = \min_j(g_{j+1}-g_j)$. Hence, if $n\geq n_0$ then
\[ |g_n-g_{n_0}| = (g_n-g_{n-1})+\dots+(g_{n_0+1}-g_{n_0}) \geq  |n-n_0|\Delta, \]
and the same relation also follows for $n\leq n_0$, implying for any probability distribution $\{p_n\}$ of $G$ that
\[ \langle |G-g_{n_0}|\rangle = \sum_n |g_n-g_{n_0}|\,p_n \geq  \sum_n |n-n_0|\Delta\,p_n =  \overline{n}\Delta, \]
 where $\overline{n}:=\langle |n-n_0|\rangle$ as before. Since inequality (\ref{number}) bounds the maximum possible entropy of any  discrete  distribution $\{p_n\}$ for a fixed value of $\overline{n}$,  it follows immediately that the entropy of $G$ is bounded by
\[ H(G) \leq  \ln [2\overline{n}+1] + 1 \leq \ln \left[ 2\Delta^{-1}\langle |G-g_{n_0}|\rangle +1\right] +1 .\]
This establishes the bound (\ref{hmax}) for an arbitrary discrete spectrum.  Again, if the spectrum of $G$ is bounded below by $g_{\rm min}$, the factor of 2 may be replaced by unity for the choice $g_{n_0}=g_{\rm min}$.

\subsection{Continuous generators}

If $G$ has a continuous spectrum, then a bound on the maximum entropy of $G$, under the constraint of a fixed value of $\langle |G- g' |\rangle$, is obtained by maximising the variational quantity
\[ J = -\int dg\,p(g)\ln p(g) + \alpha \int dg\, p(g) +\beta\int dg\,|g- g' |\,p(g) \]
over all probability distributions on the real line.  It is convenient to work with the displaced distribution $q(g):=p(g+ g' )$, having the same entropy as $p(g)$, for which the variational equation $\delta J/\delta q=0$ yields $q(g) =Ae^{-\beta|g|}$.  The constants $A$ and $\beta$ are determined by the constraints to be $A=1/(2\overline{g})$ and $\beta=1/\overline{g}$, where $\overline{g}:=\langle |G- g' |\rangle = \int dg\,|g|\,q(g)$, allowing the maximum possible entropy to be calculated as
\[ H_{\rm max}(G) = -\int dg\,q(g)[\ln A -\beta|g|] = -\ln A +\beta\overline{g} = \ln [2\overline{g}] +1, \]
establishing the bound (\ref{hmax}) for an arbitrary continuous spectrum.  If the spectrum of $G$ is bounded below by $g_{\rm min}$, a similar calculation for the choice $ g' =g_{\rm min}$ leads to $H_{\rm max}(G)\leq \ln (\overline{g}+1) +1$, i.e., to  replacement of the factor of 2 by unity. \\

\end{document}